\documentclass[prl,twocolumn,english,superscriptaddress,footinbib]{revtex4-1}

\usepackage{amsmath,amssymb,mathrsfs}
\usepackage{amsfonts,bm}
\usepackage{amsmath}
\usepackage{amssymb}
\usepackage{amsthm}
\usepackage{graphicx}
\usepackage{graphics}
\usepackage{subfigure}
\usepackage{color}
\usepackage{bm}
\usepackage{hyperref}
\usepackage{rotating}
\usepackage{comment}
\usepackage[colorinlistoftodos]{todonotes}

\usepackage{xfrac}

\newcommand{\be}{\begin{equation} }
\newcommand{\ee}{\end{equation} }
\newcommand{\ba}{\begin{eqnarray} }
\newcommand{\ea}{\end{eqnarray} }

\newcommand{\bpm}{\begin{pmatrix}}
\newcommand{\epm}{\end{pmatrix}}
\newcommand{\bmm}{\begin{matrix}}
\newcommand{\emm}{\end{matrix}}

\newcommand{\la}{\label}
\newcommand{\p}{\partial}
\newcommand{\bea}{\begin{eqnarray}}
\newcommand{\eea}{\end{eqnarray}}

\makeatother

\usepackage{babel}

\begin{document}

\title{Coastal Kelvin Mode and the Fractional Quantum Hall Edge}
 \author{Gustavo M. Monteiro}
 \affiliation{Department of Physics and Astronomy, College of Staten Island, CUNY, Staten Island, NY 10314, USA}
 \author{Sriram Ganeshan}
 \affiliation{Department of Physics, City College, City University of New York, New York, NY 10031, USA }
\affiliation{CUNY Graduate Center, New York, NY 10031}

\date{\today}

%%%%%%%%%%%%%
\begin{abstract}
 This letter explores the relationship between the coastal Kelvin mode observed in the shallow water model of ocean waves and the edge mode of a fractional quantum Hall (FQH) state. The hydrodynamic equations for the FQH state can be written as a generalized form of the shallow water equations with Coriolis force, where the density replaces the height of the fluid column and the magnetic field plays the role of the Coriolis parameter. In the FQH case, the potential vorticity associated with the shallow water model becomes a constant. In contrast to the shallow water equations, the hydro system for the FQH state contains higher derivatives of velocity which enforces the no-stress boundary condition in addition to the no-penetration condition at the hard wall or coastal boundary. For these boundary conditions, the linearized edge dynamics has two chiral edge modes propagating in the same direction: a non-dispersing Kelvin mode and a dispersing chiral boson mode. We investigate the nature of these modes in the presence of a tangent electric field. Our results show that the Kelvin mode cannot be excited by this field, and as a result, it cannot transport charge along the edge. However, the dispersive chiral boson mode is compatible with the edge dynamics of the FQH state and satisfies the anomaly equation.  \end{abstract}
%%%%%%%%%%%%%
\maketitle

%%%%%%%%%%%%%%%%%%%%%%%%
%%%%%%%%%%%%%%%%%%%%%%%%
\textit{\textbf{Introduction:}}  The Coriolis force, caused by the Earth's rotation, has a significant influence on oceanic currents and plays a crucial role in the formation of Kelvin waves. Originally discovered by Lord Kelvin~\cite{thomson18801}, these waves can occur either at the coast (coastal Kelvin waves) or near the equator (Equatorial Kelvin modes), where the Coriolis force changes direction. Recent works have noted a connection between the chiral nature of the Kelvin wave and certain topological characteristics of the linearized shallow water equations (SWE), which models the horizontal flow of a fluid layer with varying height affected by a Coriolis force \cite{delplace2017topological, tauber2019bulk}. Recent work by David Tong~\cite{tong2022gauge}  has further established that the SWE can be reformulated as a gauge theory in 2+1 dimensions featuring a Chern-Simons term. 

In this letter, we investigate how the coastal Kelvin modes are related to edge modes of fractional quantum Hall (FQH) states. Our starting point is the hydrodynamic equations obtained from the Chern-Simons-Ginzburg-Landau (CSGL) theory  which consists of a charged condensate coupled to a statistical gauge field~\cite{read1989order, zhang1989effective, zhang1992chern, stone1990superfluid, abanov2013effective, monteiro2022topological}. This FQH superfluid satisfies the SWE albeit with constant potential vorticity~\footnote{In the shallow water model, the potential vorticity is the sum of relative fluid vorticity and the Coriolis parameter divided by the ocean height. This quantity is transported by the flow. See David Tong's lecture notes on fluid mechanics for details.} and additional higher gradient terms.
 
 In contrast to the standard SWE, the presence of higher derivative terms in the FQH superfluid dynamics requires two boundary conditions for the velocity fields: no-penetration (impermeability) condition along with either the no-slip or the no-stress boundary condition for a hard wall with no dissipation. The no-slip condition means that the fluid sticks to the wall, i.e., the tangential velocity must vanish at the boundary. The no-stress condition, on the other hand, allows the fluid to slip at the wall, as long as the flow does not generate tangential forces at the boundary. We recently showed that the no-slip boundary condition forbids any edge dynamics, which is in contradiction to the FQH edge physics. However, the no-stress boundary condition gives rise to chiral edge dynamics akin to the FQH state in the presence of boundaries~\cite{monteiro2022topological}.

 %However, the no-stress boundary conditions results in a consistent anomaly equation in the presence of a tangent electric field and leads to chiral edge dynamics akin to the FQH state with boundaries.

This work shows that the FQH fluid dynamics with no-penetration and no-stress boundary conditions lead to two chiral modes propagating in the same direction.  The dispersion relation of the two edge modes is given by
 %\begin{align}
	%\omega_{K} &= c k,\\ \omega_{CB} &= 2 c k- \frac{c\ell_B^2}{2} k^3+\mathcal{O}(k^5).
%\end{align}
\be
\omega_{\text{K}} = c k,\quad \omega_{\text{CB}} = 2 c k- \frac{c\ell_B^2}{2} k^3+\mathcal{O}(k^5)\,,
\ee
where $\omega_{\text{K}, \text{CB}}$ are their frequencies, $k$ is their wavelength, $\ell_B=\sqrt{\hbar/(eB)}$ is the magnetic length and $c$ is the interaction dependent effective sound velocity. The non-dispersive mode with frequency $\omega_{\text{K}}$ has the same structure as the coastal Kelvin wave and it will be called Kelvin mode, whereas the other mode, with frequency  $\omega_{\text{CB}}$ will be referred as chiral boson mode. 

From the FQH viewpoint, the presence of two modes at the edge of a Laughlin state is somewhat puzzling since only one chiral mode is expected for states with filling factor $\nu=1/(2p+1)$, with $p\in\mathbb N$~\cite{wenbook}. In order to further investigate the nature of these modes, we turn on a tangent electric field and check the consistency of these modes with the gauge anomaly. Remarkably, we find that only the chiral boson mode is consistent with the gauge anomaly, whereas the Kelvin mode can only exist as a homogeneous solution of the corresponding hydrodynamic equations. In other words, a tangent electric field cannot excite the Kelvin mode, indicating that such a mode cannot lead to charge transport at the boundary.

The CSGL theory with hard-wall boundary has been previously considered in Refs.~\cite{nagaosa1994chern, orgad1996coulomb, orgad1997chern}.  In these papers, the linearized edge dynamics are derived in the absence of any electric field.  In Ref.~\cite{nagaosa1994chern}, the authors neglected the higher gradient quantum pressure in the Euler equation, leading to a dynamical system identical to the SWE subject to only no-penetration condition. In this case, the Kelvin mode is identified as the FQH edge mode as its inconsistency with the gauge anomaly is concealed. In Refs.~\cite{orgad1996coulomb, orgad1997chern}, while the quantum pressure is retained, the authors impose the vanishing of the condensate density at the boundary. Nevertheless, the fluid density vanishing at the wall is unusual from a superfluid point of view and is incompatible with the gauge anomaly at the edge. Thus our work emphasizes the crucial role of a tangent electric field and gauge anomaly in testing the consistency and predictions of boundary conditions in FQH fluid dynamics.

{\it \textbf{FQH hydrodynamics:}} In the saddle point approximation, the composite boson dynamics of the CSGL model can be fully described in terms of classical hydrodynamic equations with an additional constitutive relation, called Hall constraint~\cite{stone1990superfluid, abanov2013effective, monteiro2022topological}. This Hall constraint fixes the potential vorticity to be a constant, pinning the superfluid vorticity to fluctuations of the condensate density. 

Let $n$ be the condensate density and $v^i$ its velocity flow. In the presence of a strong magnetic field $B$, perpendicular to the sample, and in-plane electric field $E_i$, the FQH superfluid dynamics can be described by   
\begin{align}
	&\p_t n+\p_i(nv^i)=0\,, \la{continuity}\\
	&\p_t v_i+v^j\p_j v_i=\frac{1}{m n}\p_jT^{j}_{\,\,\,i}-\frac{e}{m}(E_i+B\epsilon_{ij} v^j),  \la{Euler}\\
&\left[\epsilon^{ij}\p_iv_j -\omega_B+\frac{\hbar}{m}\p_i\left(\frac{\p^i n}{n}\right)\right]\frac{1}{n}=-\frac{2\pi\hbar}{\nu m}\,, \la{Hall-constraint}
\end{align}
where $m$ is the effective electron mass, $\nu$ is the filling factor, $\omega_B=eB/m$ is the cyclotron frequency and $T^j\,_i$ is the stress tensor, which is given by
\begin{equation}
T^{j}_{\,\,\,i}=\Big(V-nV'(n)\Big)\,\delta^{j}_{\,\,i} -\frac{\hbar n}{2}\left(\epsilon_{ik}\p^kv^j+\epsilon^{jk}\p_iv_k\right). \la{stress}
\end{equation}
The first term in Eq.~(\ref{stress}) is the fluid pressure with an overall negative sign, whereas the second one is the odd viscosity term~\cite{avron1995viscosity, avron1998odd, ganeshan2017odd, abanov2020hydrodynamics}. In two dimensions, the quantum pressure coming from the condensate dynamics can be traded for an odd viscosity term by using a particular velocity redefinition~\cite{geracie2015hydrodynamics, monteiro2021hamiltonian}. This redefinition does not modify the continuity equation but gives rise to the odd viscosity term and a second-order derivative modification to the potential vorticity. The Hall constraint~(\ref{Hall-constraint}) simply states that the potential vorticity of the FQH fluid is constant and equal to $-2\pi \hbar/(\nu m)$.

The set of equations~(\ref{continuity}-\ref{Euler}, \ref{stress}) is commonly referred to as first-order hydrodynamic equations since the stress tensor contains up to first-order gradient terms. In contrast, the  SWE are known as perfect fluid equations and does not contain gradient corrections to the stress tensor. 

The electron-electron interaction plays the role of the internal energy of the fluid. When the electron-electron interaction is sufficiently shielded in the sample, $V(n)$ can be approximated by a local two-body potential. Throughout this work, we assume that the function $V(n)$ can be described by Mexican hat potential, that is, $V(n)=\lambda\left(n-\frac{\nu e B}{2\pi \hbar}\right)^2$.

{\it \textbf{Hard wall boundary conditions:}}   For simplicity, let us assume that the FQH superfluid permeates the whole lower half-plane and that there is a hard wall at $y=0$. The first-order hydrodynamics requires two boundary conditions for the velocity fields: no-penetration condition along with either a no-slip or a no-stress boundary condition for a hard wall. For FQH fluids, the no penetration condition is modified in the presence of a tangent electric field due to an anomaly inflow mechanism. That is, the tangent electric field drives a normal current into the boundary, leading to
\begin{align}
	\left( n v_y+\frac{e\nu}{2\pi\hbar} E_x\right)\bigg|_{y=0} =0\,. \la{anomaly-inflow}
\end{align}
This charge is then forced to flow along the edge by the same tangential electric field which is in contradiction with the no-slip condition.

On the other hand, the no-stress condition gets modified at the edge due to the gauge anomaly~\cite{monteiro2022topological}, resulting in 
\begin{align}
	 T_{yx}\Big|_{y=0}&=-\frac{\nu e}{4\pi}\Big(\sqrt{ n}+\frac{\p_y  n}{ n}\Big) E_x\Big|_{y=0}\,. \la{no-stress}
\end{align}
This equation can also be written as a dynamical equation for boundary fields, which has the form of the gauge anomaly equation,
\begin{align}
\left[\p_t(\sqrt{ n})+\p_x(\sqrt{ n}\, v_x)+\frac{\nu e}{4\pi\hbar} E_x\right]\bigg|_{y=0}&=0\,. \la{edge-continuity}	
\end{align}
It is worth pointing out that Eqs.~(\ref{anomaly-inflow}-\ref{no-stress}) reduce to no-penetration and no-stress conditions respectively in the absence of an electric field.

{\it \textbf{Linearized dynamics:}} We now investigate the mode structure of the Eqs.~(\ref{continuity}-\ref{stress}) subject to the boundary conditions~(\ref{anomaly-inflow}-\ref{no-stress}). For that, we consider small perturbations on top of a constant and uniform background solution. Since we are interested in linear response theory, the electric field is also assumed to be small. Denoting $n=\frac{\nu e B}{2\pi \hbar}(1+\rho)$ and linearizing the bulk equations, we obtain
\begin{align}
	&\p_t \rho+\p_iv^i=0\,,\label{eq:linearbulk1}\\
	&\p_tv_i+c^2\p_i\rho+\omega_B\left(1+\frac{\ell_B^2}{2}\Delta\right)\epsilon_{ij}v^j=-\frac{e}{m}E_i\,,\label{eq:linearbulk2}\\
	&\left(1+\frac{\ell_B^2}{2}\Delta\right)\rho+\frac{1}{\omega_B}\epsilon_{ij}\p_iv_j=0\,, \label{eq:linearbulk3}
\end{align}
where $\Delta=\p_x^2+\p_y^2$ denotes the 2D Laplacian operator and the sound velocity $c$ is defined by the interaction strength, that is, $c^2=\omega_B\nu\lambda/(2\pi\hbar)$. 

The linear form of the boundary conditions~(\ref{anomaly-inflow}, \ref{edge-continuity}) can be written as
\begin{align}
	\left(v_y+\frac{E_x}{B}\right)\bigg|_{y=0}&=0\,,\label{eq:linearbc1}
 \\ 	\left(\p_t\rho+2\,\p_xv_x+\sqrt{\frac{\nu}{2\pi}}\frac{E_x}{B\ell_B}\right)\bigg|_{y=0}&=0\,.\label{eq:linearbc2}
\end{align}

For constant and uniform electric fields, we can absorb the inhomogeneous source terms in Eqs.~(\ref{eq:linearbulk2}-\ref{eq:linearbc1}) into background solutions by using 
\be
v_x=\frac{E_y}{B}+u\,,\qquad v_y=-\frac{E_x}{B}+v\,.
\ee
Therefore, we can set $E_i=0$ at Eqs.~(\ref{eq:linearbulk1}-\ref{eq:linearbc1}), if we replace $v_x\rightarrow u$ and $v_y\rightarrow v$ in the same equations. However, the boundary condition~(\ref{eq:linearbc2}) still remains an inhomogeneous equation, given by
\be
\left(\p_t\rho+2\,\p_xu+\sqrt{\frac{\nu}{2\pi}}\frac{E_x}{B\ell_B}\right)\bigg|_{y=0}=0\,.\label{eq:linearbc2u}
\ee

{\it \textbf{SWE and Coastal Kelvin waves:}} In Ref.~\cite{nagaosa1994chern}, the authors neglected the terms containing the Laplacian operator in Eqs.~(\ref{eq:linearbulk2}-\ref{eq:linearbulk3}). Even though this is often justified in the hydrodynamic context, where higher-order gradients are regulated by the mean free path of the fluid particles, this is not the case for the FQH superfluid. We showed in~\cite{monteiro2022topological} that these higher-order derivatives become important near the boundary, leading to a compressible boundary layer with non-trivial chiral edge dynamics. 

 In fact, the hydrodynamic equations considered by Nagaosa and Kohmoto in Ref.~\cite{nagaosa1994chern} are exactly the linearized SWE with constant potential vorticity. If we relax the condition of constant potential vorticity, these equations are known to describe the dynamics of ocean waves at high latitudes. For wavelengths much smaller than the Earth's circumference and much larger than the height of ocean waves, we can neglect the Earth's curvature and approximate the dynamics as
\begin{align}
    &\p_t h+H\p_iv^i=0\,, \la{h-shallow}
\\
&\p_t v_i+g\p_ih-f\epsilon_{ij} v^j=0\,, \la{v-shallow}
\end{align}
where $h$ the ocean wave height, $H$ is the average ocean depth, $g$ is the acceleration due to gravity and $f$ is the Coriolis parameter. In terms of the latitude $\vartheta$, the Coriolis parameter is given by $f=4\pi\sin\vartheta/\text{day}$. 

Comparing Eqs.~(\ref{h-shallow}-\ref{v-shallow}) with Eqs.~(\ref{eq:linearbulk1}-\ref{eq:linearbulk2}), we see that $h/H$ plays the role of the density fluctuation $\rho$ and $\sqrt{gH}$ is identified with the speed of sound $c$.  

The presence of a continental coast imposes the no-penetration boundary condition 
\be
v_y\big|_{y=0}=0\,, 
\ee
which allows for exponentially localized solutions near the coast at $y=0$. These solutions are called coastal Kelvin waves and satisfy the following unidirectional dispersion relation
\be
\omega_{\text{K}}=-\text{sgn}(f)\sqrt{gH} \,k\,,
\ee
as well as $v_y=0$ everywhere and constant potential vorticity. In the following, we show that the Kelvin mode is also present in the FQH case, but does not lead to the chiral edge transport in the presence of tangent electric field. 

{\it \textbf{FQH edge modes:}} Without boundaries we can look for plane wave solutions, $(\rho, u, v) \propto e^{i\bm{q}\cdot\bm{x}-i\omega t}$, for Eqs.~(\ref{eq:linearbulk1}-\ref{eq:linearbulk3}).  This gives us the following dispersion for the bulk magnetoplasmons  
\begin{align}
	\omega=\pm \sqrt{c^2q^2+\omega_B^2\left(1-\tfrac{1}{2}q^2\ell_B^2\right)^2}\,.
 \label{eq:disp-bulk}
\end{align}
In the FQH case, the smallest magnetoplasmon band gap happens at $q=0$, which gives us $c^2\geq \ell_B^2\omega_B^2$ and consequently constrains the value of the coupling parameter to be $\lambda\geq \hbar\omega_B(2\pi\ell_B^2/\nu)$.

 For the FQH superfluid confined to the lower half-plane $y\leq 0$, we must look for solutions of the type $e^{s y +i kx -i\omega t}$, with $\operatorname{Re}(s)>0$. These edge waves can be obtained from the bulk propagating solutions through the substitution $(q_x,q_y)\to (k,-is)$. Thus, Eq.~(\ref{eq:disp-bulk}) allows us to determine $s$ for a fixed frequency $\omega$ and a fixed wavenumber $k$. It is easy to see that Eq.~(\ref{eq:disp-bulk}) can be recast as a quartic degree polynomial in $s$, with roots $(s_1,s_2,-s_1,-s_2)$. Among them, only two of the roots satisfy $\operatorname{Re}(s)>0$, which we denote by $s_1$ and $s_2$. For $c^2\geq 2\ell_B^2\omega_B^2$, $s_1$ and $s_2$ become real, which indicates inter-band mixing. To avoid that, we must impose that the interaction strength satisfies $1\leq \nu\lambda/(\pi\ell_B^2\hbar\omega_B)<2$ with $s_2=s_1^*$. Following Lamb \cite{lamb1932hydrodynamics}, we look for the solution as a linear superposition of these two boundary waves
\begin{align}
	\begin{pmatrix}
\rho \\
u \\
v
\end{pmatrix} &= e^{ikx-i\omega t}\sum^2_{\alpha=1}
	\frac{ C_\alpha e^{s_\alpha y}}{k^2-s_\alpha^2} \begin{pmatrix}
k^2-s_\alpha^2 \\
\omega k -\omega_B Q_{\alpha} s_\alpha \\
i\omega_B Q_{\alpha} k-i\omega s_\alpha 
\end{pmatrix} ,
 \label{eq:vector}
\end{align}
where $Q_{\alpha}:=1+\frac{1}{2}(s_{\alpha}^2-k^2)\ell_B^2$ and both eigenvectors at the RHS are solutions to the linearized bulk equations. 

Imposing $v(x,0,t)=0$ splits the boundary solutions into two distinct cases,  that is, $\omega=ck$ (Kelvin mode) and $\omega\neq ck$ (Chiral Boson mode). 

{\it \textbf{Kelvin mode with $E_x=0$:}} For the Kelvin mode case, simply imposing $\omega=ck$ forces $v$ to vanish everywhere, leading to 
\be
\begin{pmatrix}
\rho_\text{K} \\
u_\text{K} \\
v_\text{K}
\end{pmatrix} = \begin{pmatrix}
1 \\
c \\
0 
\end{pmatrix}e^{ik(x-c t)}\sum^2_{\alpha=1}
	 C_\alpha e^{s_\alpha y}  \,, \la{eq:kelvin}
\ee
together with
\be
s_1=s_2^*=\frac{c}{\omega_B\ell_B^2}+\frac{i}{\ell_B}\sqrt{2-\frac{c^2}{\ell_B^2\omega_B^2}-k^2\ell_B^2}\,.
\ee
It is interesting to note that both the dispersion ($\omega=c k$) and the eigenvector ($v_\text{K}=0$ and $u_\text{K}=c \rho_\text{K}$) are fully determined within the homogeneous equations, i.e., bulk equations together with the no penetration condition. This dispersion and  the eigenvector have the same form as in the case of the coastal Kelvin mode. Furthermore, the non-dispersive nature of the Kelvin mode enforces a strong constraint on the charge density, that is, $\rho_\text{K}(x,y,t)=\rho_\text{K}(x-ct,y)$. This constraint has drastic consequences for the consistency with the inhomogeneous boundary condition given in Eq.~(\ref{eq:linearbc2u}) as we will show in subsequent sections with the electric field. The boundary condition in Eq.~(\ref{eq:linearbc2u}) then gives us $C_1=-C_2$ for the Kelvin mode, forcing  density fluctuations to vanish at the boundary for $k\neq 0$ and to be constant for $k=0$. In fact, the lack of edge dynamics for the Kelvin mode holds not only for Eq.~(\ref{eq:linearbc2u}), but for any non-trivial second boundary condition.
%\be
%\begin{pmatrix}
%\rho_K \\
%u_K \\
%v_K
%\end{pmatrix} &= C\begin{pmatrix}
%1 \\
%c \\
%0 
%\end{pmatrix}e^{\operatorname{Re}(s_1) \,y+ik(x-c t)}\sin\Big(\operatorname{Im}(s_1)\, y}\Big)\,.
%\ee

\begin{figure}
\centering
\includegraphics[scale=0.2]{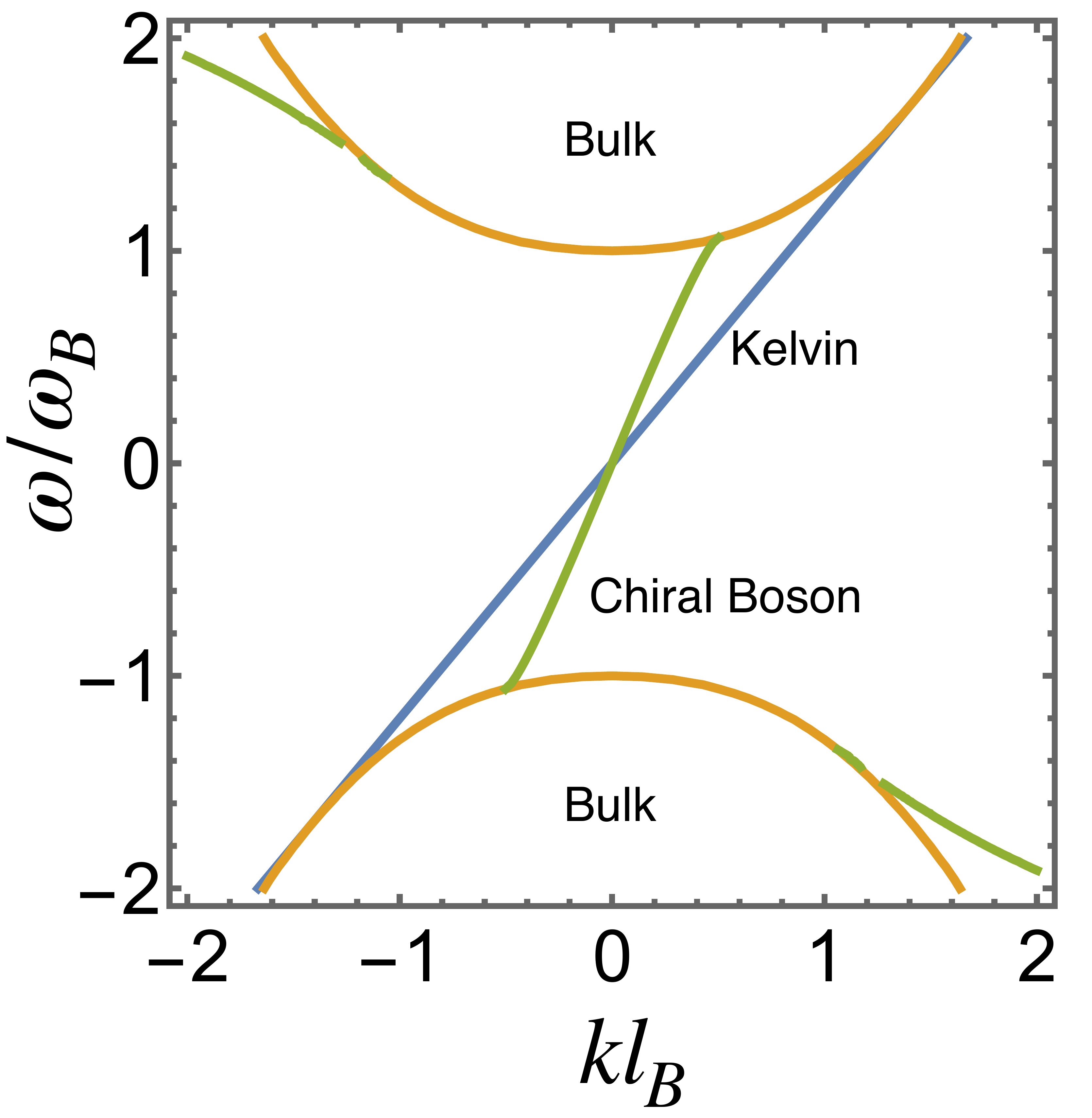}
\caption{Numerical dispersion relation for the bulk and edge modes for the FQH hydro subject two no-penetration and no-stress boundary condition. We set $c=1.11 \ell_B\omega_B$.}
\label{fig:disp}
\end{figure}

{\it \textbf{Chiral boson mode with $E_x=0$:}} We now consider the $\omega\neq ck$ case and impose the no-penetration condition $v(x,0,t)=0$. This boundary condition is satisfied when
 \begin{align}
 	C_1=-C_2^*=C\,\frac{\omega s_1+\omega_B k Q_1}{\omega^2-c^2k^2}\,,
 	\label{eq:sol2}
 \end{align}
with $C\in\mathbb R$. Note that, unlike in the Kelvin mode case, this boundary condition alone does not determine the dispersion relation. For that, we need the second boundary condition Eq.~(\ref{eq:linearbc2u}), with $E_x=0$, which results in a transcendental equation. For calculation details we refer to the Appendix. For small frequencies and long wavelengths, this chiral boson dispersion relation can be expressed as a series in powers of $k$
\be
\omega_{\text{CB}}=2ck-\frac{c\ell^2_B}{2}k^3+\mathcal O(k^5)\,.
\ee
In contrast to the Kelvin mode, the chiral boson mode is not usually seen as the coastal mode, except in Ref.~\cite{tauber2019bulk} where odd viscosity was used as a regulator and was retained as an additional no stress boundary condition. In Fig.~\ref{fig:disp}, we see the non-dispersive Kelvin mode and the dispersive chiral boson mode together with the gapped bulk spectrum. In Fig.~\ref{fig:disp}, there are additional edge modes emerging from the bulk at energies above the gap scale which is also noted in \cite{tauber2019bulk} for the SWE. 
%%%%%%%%%%%%%%%%%%%%%%%%%
%%%%%%%%%%%%%%%%%%%%%%%%%%

%Similar mode structure has been reported in the coastal Kelvin wave problem with an odd viscosity induced no-stress boundary condition. The odd viscosity in the geophysical shallow water dynamics is used as a regulator and is taken to zero at the end, but in the FQH case this term is tied to the quantum-pressure term and cannot be dropped. The presence of two modes at the edge of a Laughlin FQH state  is puzzling since we expect only one chiral mode corresponding to $\nu=1/{2m+1}$ filling factors. 

  {\it \textbf{Gauge anomaly:}} To investigate the compatibility of these chiral modes with the FQH edge dynamics, we turn on a constant tangent electric field $E_x$. The characteristics of these modes can be determined by examining the dynamical boundary condition Eq.~(\ref{eq:linearbc2u}) for the charge density at the edge.  We determine $u(x,0,t)$ in terms of boundary density $\rho(x,0,t)$ and its derivatives by solving Eqs.~(\ref{eq:linearbulk1}-\ref{eq:linearbc1}) and express the inhomogeneous partial differential equation~(\ref{eq:linearbc2u}) solely in terms of boundary density and the electric field. This is equivalent to integrating out the bulk velocity fields, ending up with effective 1D dynamics for the edge density in the presence of the electric field.  We again separate this analysis into two cases, namely, the Kelvin mode and the chiral boson mode.  
 
{\it \textbf{Kelvin mode with $E_x\neq 0$:}} We can see from Eq.~(\ref{eq:kelvin}) that for the Kelvin mode we have
 \be
u_\text{K}(x,y,t)=c \rho_\text{K}(x,y,t)=c \rho_\text{K}(x-ct,y)\,. \la{u-rho-kelvin}
 \ee
Plugging the solution~(\ref{u-rho-kelvin}) into Eq.~(\ref{eq:linearbc2u}), we find that
\begin{align}
\rho_\text{K}(x-ct,0)&=\sqrt{\frac{\nu}{2\pi}}\frac{E_x}{B\ell_B}\left(t-\frac{x}{c}\right)+\rho_0\,,
\end{align}
where $\rho_0$ is a constant. This expression for the boundary density is unphysical since the boundary charge is unbounded at infinity for any non-zero value of the electric field. This means that the Kelvin mode cannot transport charge along the edge in the presence of tangent electric field and, consequently, fails to capture the FQH edge physics.

{\it \textbf{Chiral boson mode with $E_x\neq 0$:}} For this mode, the relation between $u_{\text{CB}}(x,0,t)$ and $\rho_{\text{CB}}(x,0,t)$ is more intricate. However, using Eqs.~(\ref{eq:vector}) and (\ref{eq:sol2}), we can expand $u_{\text{CB}}(x,0,t)$ in terms of $\rho_{\text{CB}}(x,0,t)$ and its spatial derivatives in the small $k$ limit.  The first two leading terms of this expansion are given by 
  \begin{align}
 	u_{\text{CB}}(x,0,t)=c\left(1+\frac{\ell_B^2}{4}\p_x^2+\mathcal{O}(\p_x^4)\right)\rho_{\text{CB}}(x,0,t)\label{eq:gradexp}
 \end{align}
 Substituting this expression into Eq.~(\ref{eq:linearbc2u}) and truncating the derivative beyond $\mathcal{O}(\p_x^4)$,  we obtain    
  \begin{align}
 	&\rho_{\text{CB}}(x,0,t)=-\sqrt{\frac{\nu}{2\pi}}\frac{E_x}{B\ell_B}t+\eta(x,t)\,,
 	\end{align}
 	with
 	\begin{align}
 	&\left(\p_t+2c \,\p_x+\frac{c \ell_B^2}{2}\p_x^3\right)\eta=0\,.\label{eq:hom2}
 \end{align}
 The boundary charge dynamics associated with the chiral boson mode does not have any issues with the anomaly inflow mechanism, since it is bounded at infinity. 
Therefore, the general solution to the full inhomogeneous system contains both the Kelvin and the chiral boson modes, however, the former can only appear as a homogeneous solution (which has a constant value at the boundary and is not modified by the electric fields) and hence plays no role in the boundary charge transport.

{\it \textbf{Discussion:}} This letter has investigated the relationship between the coastal Kelvin mode in a shallow water model with Coriolis force and the edge mode of a fractional quantum Hall (FQH) state. The FQH fluid dynamic equations on the half-plane have two chiral edge modes that move in the same direction: a non-dispersive Kelvin mode and a dispersing chiral boson mode. Our work has shown that the Kelvin mode is incompatible with the gauge anomaly and, thus, cannot be associated with the charge transport at the edge. Additionally, from the dispersion relation in Fig.~\ref{fig:disp}, one can see that the Kelvin mode is tangent to bulk dispersion and does not merge into the bulk spectrum. However, typically in Integer quantum Hall models with bulk Hamiltonian (such as the Hofstadter model), the edge mode necessarily disperses to merge with the bulk bands. 

Past works on the CSGL model with boundaries did not consider tangent electric fields, which overlooked the Kelvin mode's incompatibility with the FQH edge anomalous dynamics. The chiral boson mode, on the other hand, is consistent with the FQH edge dynamics. 

Our work suggests two potential conclusions regarding the presence of these modes at the edge of the FQH sample. The first possibility is that the Kelvin mode may be part of the FQH edge dynamics, but not detectable in transport experiments since it cannot be excited by electromagnetic fields. Alternatively, the Kelvin mode may be an artifact of our framework and perhaps needs to be eliminated for an accurate description of the FQH state. To investigate this further, future works should delve into heat transport, non-linear corrections, and the role of symmetries like Galilean invariance in relation to the Kelvin mode.

 \textit{\textbf{Acknowledgments.}} We would like to thank V. P. Nair, David Tong, T. H. Hansson, and Alexander Abanov for stimulating discussions. GMM would also like to thank D. X. Nguyen and D. T. Son for fruitful discussions. SG is supported by NSF CAREER Grant No. DMR-1944967 (SG).  Part of this work was performed at the Aspen Center for Physics, which is supported by National Science Foundation grant PHY-1607611. ported by NSF CAREER Grant No. DMR-1944967 (SG). GMM was supported in part by the National Science Foundation under Grant OMA1936351.
 
\bibliographystyle{my-refs}
\bibliography{oddviscosity-bibliography.bib}

\begin{thebibliography}{10}

\bibitem{thomson18801}
W.~Thomson.
\newblock \emph{1. On gravitational oscillations of rotating water}.
\newblock Proceedings of the Royal Society of Edinburgh, \textbf{10}, 92--100
  (1880).

\bibitem{delplace2017topological}
P.~Delplace, J.~Marston, and A.~Venaille.
\newblock \emph{Topological origin of equatorial waves}.
\newblock Science, \textbf{358}, 1075--1077 (2017).

\bibitem{tauber2019bulk}
C.~Tauber, P.~Delplace, and A.~Venaille.
\newblock \emph{A bulk-interface correspondence for equatorial waves}.
\newblock Journal of Fluid Mechanics, \textbf{868}, R2 (2019).

\bibitem{tong2022gauge}
D.~Tong.
\newblock \emph{A Gauge Theory for Shallow Water}.
\newblock arXiv preprint arXiv:2209.10574 (2022).

\bibitem{read1989order}
N.~Read.
\newblock \emph{Order parameter and Ginzburg-Landau theory for the fractional
  quantum Hall effect}.
\newblock Physical Review Letters, \textbf{62}, 86 (1989).

\bibitem{zhang1989effective}
S.~C. Zhang, T.~H. Hansson, and S.~Kivelson.
\newblock \emph{Effective-field-theory model for the fractional quantum Hall
  effect}.
\newblock Physical review letters, \textbf{62}, 82 (1989).

\bibitem{zhang1992chern}
S.~C. Zhang.
\newblock \emph{The Chern--Simons--Landau--Ginzburg theory of the fractional
  quantum Hall effect}.
\newblock International Journal of Modern Physics B, \textbf{6}, 25--58 (1992).

\bibitem{stone1990superfluid}
M.~Stone.
\newblock \emph{Superfluid dynamics of the fractional quantum Hall state}.
\newblock Physical Review B, \textbf{42}, 212 (1990).

\bibitem{abanov2013effective}
A.~G. Abanov.
\newblock \emph{On the effective hydrodynamics of the fractional quantum Hall
  effect}.
\newblock Journal of Physics A: Mathematical and Theoretical, \textbf{46},
  292001 (2013).

\bibitem{monteiro2022topological}
G.~M. Monteiro, V.~Nair, and S.~Ganeshan.
\newblock \emph{Topological fluids and FQH edge dynamics}.
\newblock arXiv preprint arXiv:2203.06516 (2022).

\bibitem{Note1}
In the shallow water model, the potential vorticity is the sum of relative
  fluid vorticity and the Coriolis parameter divided by the ocean height. This
  quantity is transported by the flow. See David Tong's lecture notes on fluid
  mechanics for details.

\bibitem{wenbook}
X.-G. Wen.
\newblock \emph{Quantum field theory of many-body systems: from the origin of
  sound to an origin of light and electrons.}
\newblock Oxford University Press (2004).

\bibitem{nagaosa1994chern}
N.~Nagaosa and M.~Kohmoto.
\newblock \emph{Chern-Simons Ginzburg-Landau Theory of the Fractional Quantum
  Hall System with Edges}.
\newblock In \emph{Correlation Effects in Low-Dimensional Electron Systems},
  pages 168--174. Springer (1994).

\bibitem{orgad1996coulomb}
D.~Orgad and S.~Levit.
\newblock \emph{Coulomb drag of edge excitations in the Chern-Simons theory of
  the fractional quantum Hall effect}.
\newblock Physical Review B, \textbf{53}, 7964 (1996).

\bibitem{orgad1997chern}
D.~Orgad.
\newblock \emph{From the Chern-Simons theory for the fractional quantum Hall
  effect to the Tomonaga-Luttinger model of its edges}.
\newblock Physical review letters, \textbf{79}, 475 (1997).

\bibitem{avron1995viscosity}
J.~Avron, R.~Seiler, and P.~G. Zograf.
\newblock \emph{Viscosity of quantum Hall fluids}.
\newblock Physical review letters, \textbf{75}, 697 (1995).

\bibitem{avron1998odd}
J.~Avron.
\newblock \emph{Odd viscosity}.
\newblock Journal of statistical physics, \textbf{92}, 543--557 (1998).

\bibitem{ganeshan2017odd}
S.~Ganeshan and A.~G. Abanov.
\newblock \emph{Odd viscosity in two-dimensional incompressible fluids}.
\newblock Physical Review Fluids, \textbf{2}, 094101 (2017).

\bibitem{abanov2020hydrodynamics}
A.~G. Abanov, T.~Can, S.~Ganeshan, and G.~M. Monteiro.
\newblock \emph{Hydrodynamics of two-dimensional compressible fluid with broken
  parity: variational principle and free surface dynamics in the absence of
  dissipation}.
\newblock Physical Review Fluids, \textbf{5}, 104802 (2020).

\bibitem{geracie2015hydrodynamics}
M.~Geracie and D.~T. Son.
\newblock \emph{Hydrodynamics on the lowest Landau level}.
\newblock Journal of High Energy Physics, \textbf{2015}, 1--27 (2015).

\bibitem{monteiro2021hamiltonian}
G.~M. Monteiro, A.~G. Abanov, and S.~Ganeshan.
\newblock \emph{Hamiltonian structure of 2D fluid dynamics with broken parity}.
\newblock arXiv preprint arXiv:2105.01655 (2021).

\bibitem{lamb1932hydrodynamics}
H.~Lamb.
\newblock \emph{Hydrodynamics}.
\newblock Cambridge university press (1932).

\end{thebibliography}

 \newpage
 
\onecolumngrid
\appendix
\section{Appendix}

%%%%%%%%%%%%%%%%%%%%%%%%%%

\section{Derivation of the chiral boson mode dispersion}
\label{App}
In this section, we provide the calculation details for the chiral boson mode solution. For the sake of brevity, we will use dimensionless variables, that is, $\omega/\omega_B \rightarrow \omega$, and  $c/(\omega_B \ell_B)\rightarrow c$, and $k \ell_B \rightarrow k$. The linearized bulk equations are given by
\begin{align}
	&\p_t \rho+\p_xu+\p_y v=0,\label{eq:linearbulks1}\\
	&\p_t u+c^2\p_x\rho+\left(1+\frac{1}{2}(\p_x^2+\p_y^2)\right)v=0\label{eq:linearbulks2}\\
		&\p_tv+c^2\p_y\rho-\left(1+\frac{1}{2}(\p_x^2+\p_y^2)\right)u=0\label{eq:linearbulks3}\\
	&(\p_x v-\p_y u)+\left(1+\frac{1}{2}(\p_x^2+\p_y^2)\right)\rho=0, \label{eq:constraint}
\end{align}
whereas the linearized boundary conditions can be written as
\begin{align}
	v\big|_{y=0}=0,\label{eq:linearbcs1}\\ 
	\left(\p_t\rho+2\,\p_xu+\sqrt{\frac{\nu}{2\pi}}\frac{E_x}{B}\right)\bigg|_{y=0}=0.\label{eq:linearbcs2}
\end{align}

The general solution is a linear superposition of two boundary waves
\begin{align}
	(\rho, u, v) = e^{ikx-i\omega t} \sum^2_{\alpha=1}
	C_\alpha e^{s_\alpha y}\; \left(1, \,\,\frac{\omega k -Q_{\alpha} s_\alpha}{k^2-s_\alpha^2},\,\, 
	-i\frac{\omega s_\alpha- Q_{\alpha} k}{k^2-s_\alpha^2}  \right) \,,
 \label{eq:vector-s}
\end{align}
where $s_\alpha$ are roots of following polynomial 
\begin{align}
	c^2(k^2-s_\alpha^2)+\left(1+\tfrac{1}{2}(s_\alpha^2-k^2)\right)^2-\omega^ 2=0\,.
 \label{eq:disp-bulk-s}
\end{align}
Note that, Eq.~(\ref{eq:disp-bulk-s}) can be expressed as
\be
\left(\tfrac{1}{2}(s_\alpha^2-k^2)^2-c^2+1\right)^2+c^2(2-c^2)-\omega^2=0\,,
\ee
which admits no real solutions when $c^2<1+\sqrt{1-\omega^2}$.

Imposing $v(x,0,t)=0$ onto Eq.~(\ref{eq:vector-s}) we get
\begin{align}
	C_1=\frac{C(k^2-s^2_1)}{\omega s_1- k\left(1+\frac{1}{2}(s^2_1-k^2)\right)},\quad C_2=-C_1^*, \quad\text{with}\quad C\in \mathbb{R},\label{eq:sol2-s}\end{align}
together with
\begin{align}
s_1=s_2^*=\sqrt{c^2-1+\frac{k^2}{2}+\sqrt{\left(1-\frac{k^2}{2}\right)^2+c^2 k^2-\omega^2}}+i\sqrt{1-c^2-\frac{k^2}{2}+\sqrt{\left(1-\frac{k^2}{2}\right)^2+c^2 k^2-\omega^2}}\,.
\end{align}

The next step in finding the chiral boson dispersion relation is to write $u(x,0,t)$ in terms of $\rho(x,0,t)$. Using Eq.~(\ref{eq:vector-s}) and Eq.~(\ref{eq:sol2-s}), we find that
\begin{align}
	\tilde u(k,\omega)\Big|_{y=0}= c\chi(k,\omega)\tilde \rho(k,\omega)\Big|_{y=0}, \quad \text{with}\quad \chi(k,\omega)=\left(\frac{c^2 k+ \omega\operatorname{Re}(s_1)}{c\omega+ck \operatorname{Re}(s_1)}\right)
\end{align}
In the absence of the electric field, we can write a transcendental equation that determines the dispersion relation,
\begin{align}
	\left( \omega -2 c k\chi(k,\omega)\right) \tilde\rho(k,\omega)\Big|_{y=0}=0\quad \Rightarrow \quad \omega -2 c k\chi(k,\omega) =0.
\end{align}

The full numerical chiral boson dispersion associated with this mode is shown in Fig.~\ref{fig:disp}. We can also solve the above dispersion by expanding $\omega$ in powers of $k$. The first two leading terms can be written as,
\begin{align}
	\omega_{\text{CB}}=2c k-\frac{c}{2}k^3+\mathcal{O}(k^5)
\end{align}

For the case when $E_x\neq0$, it is convenient to expand the kernel $\chi(k,\omega)$ in powers of $\omega$ and $k$, which leads to 
\begin{align}
	\tilde u(k,\omega)\Big|_{y=0}= c\left(1-\frac{k^2}{4}-\frac{\omega(\omega-2c k) }{4c^2}+....\right)\tilde \rho(k,\omega)\Big|_{y=0}.
\end{align}
The above expansion can then be rewritten as gradient expansion of spatial derivatives and after restoring the dimensions is equal to Eq.~\ref{eq:gradexp} given in the main text.

%%%%%%%%%%%%%%%%%%%%%%%%
%%%%%%%%%%%%%%%%%%%%%%%%

%%%%%%%%%%%%%%%%%%%%%%%%
%%%%%%%%%%%%%%%%%%%%%%%%

\end{document}